\documentclass[onecolumn]{aa}
\usepackage{graphicx}
\usepackage[T1]{fontenc}
\usepackage{ae,aecompl}
\usepackage{txfonts}
\begin{document}
   \title{EIT and TRACE responses to flare plasma}


   \author{D. Tripathi\inst{1}, G. Del Zanna\inst{2},
   H. E. Mason\inst{1} and C. Chifor\inst{1}}

   \offprints{D.Tripathi@damtp.cam.ac.uk}

   \institute{Department of Applied Maths and Theoretical Physics,
   University of Cambridge, Cambridge CB3 0WA, UK\\
\email{[D.Tripathi; H.E.Mason; C. Chifor]@damtp.cam.ac.uk}
\and University College London, MSSL, Holmbury St. Mary Dorking Surrey
RH5 6NT, UK\\
\email{gd232@mssl.ucl.ac.uk}}

\date{Date: Submitted, Date: Accepted}

\abstract{} {To understand the contribution of active region and flare
plasmas to the $\lambda$195 channels of SOHO/EIT (Extreme-ultraviolet
Imaging Telescope) and TRACE (Transition Region and Coronal
Explorer).} {We have analysed an M8 flare simultaneously observed by
the Coronal Diagnostic Spectrometer (CDS), EIT, TRACE and RHESSI. We
obtained synthetic spectra for the flaring region and an outer region
using the differential emission measures (DEM) of emitting plasma
based on CDS and RHESSI observations and the CHIANTI atomic database.
We then predicted the EIT and TRACE count rates.}  {For the flaring
region, both EIT and TRACE images taken through the $\lambda$195
filter are dominated by Fe ${\rm XXIV}$ (formed at about 20 MK).
However, in the outer region, the emission was primarily due to the Fe
${\rm XII}$, with substantial contributions from other lines. The
average count rate for the outer region was within 25\% the observed
value for EIT, while for TRACE it was a factor of two higher.  For the
flare region, the predicted count rate was a factor of two (in case of
EIT) and a factor of three (in case of TRACE) higher than the actual
count rate.}  {During a solar flare, both TRACE and EIT $\lambda$195
channels are found to be dominated by Fe ${\rm XXIV}$
emission. Reasonable agreement between predictions and observations is
found, however some discrepancies need to be further investigated.}

\keywords{Sun: corona - Sun: spectroscopy - Sun: flares - Sun:
emission line}
\titlerunning{Forward modelling} 
\authorrunning{D. Tripathi et al.} 

\maketitle
\section{Introduction}

The Extreme-ultraviolet Imaging Telescopes (EIT; Delaboudini\`ere et
al. 1995) and the Coronal Diagnostic Spectrometer (CDS; Harrison et
al. 1995) aboard the Solar and Heliospheric Observatory (SOHO) and the
Transition Region and Coronal Explorer (TRACE; Handy et al. 1998) have
provided a wealth of information about the different, highly dynamic
layers of solar atmosphere.  In order to understand the physics, it is
mandatory to understand the properties of plasma which is being
observed.  Multi-wavelength observations from different instruments
together with the CHIANTI atomic database (Dere et al. 1997; v.5;
Landi et al. 2006) are routinely used to study a variety of different
solar features.

The EIT provides full disk observations of the Sun in four spectral
bands, one of which is centered around $\lambda$195.  In normal quiet
Sun conditions, this band is dominated by Fe{\rm XII} emission, while
in coronal hole plumes it is dominated by Fe{\rm VIII} (Del Zanna et
al. 2003).  It has been suggested (see, e.g., McKenzie 2000) that this
channel may be contaminated by high-temperature Fe {\rm XXIV} (20 MK)
emission. However this has not been throughly studied or definitively
proven.

TRACE also provides observations of the solar corona in three coronal
band, one of which is also centered around $\lambda$195. It has also
been suggested that the TRACE $\lambda$195 channel has a high
temperature response due to Fe XXIV during flares (Warren et al. 1999,
Warren \& Reeves 2001, Gallegher et al. 2002, Phillips et
al. 2005). However no attemp has been made for a direct and
quantitative comparison.

Both EIT and TRACE have provided many images of flares, and it is
therefore important to establish the Fe XXIV contribution to this
band.  This is not a trivial task, since simultaneous multi-wavelength
observations of flares are a rare event. In this paper, we have
combined simultaneous observations of an M8 flare obtained with CDS,
RHESSI, EIT and TRACE $\lambda$195 images. We have derived the
differential emission measure (DEM) from CDS and RHESSI data for a
flaring region and an outer region and then performed forward
modelling with the CHIANTI atomic database to predict the EIT and
TRACE count rates.

\section{Observations and Data Reduction}

\begin{figure}[!b]
\centering
\includegraphics[width=0.3\textwidth]{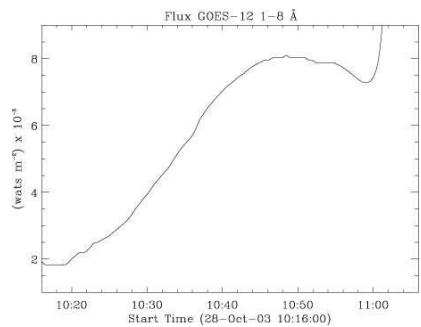}
\caption{The GOES-12 plot for the flare in the 
1.0-8.0 {\AA} channel. \label{goes}}
\end{figure}

We performed a search through the entire CDS NIS (Normal Incidence
Spectrometer) database to find suitable flare events. Out of the many
flare datasets recorded by CDS, only an M8-class flare on 28-Oct-2003
at around 10:48 UT was simultaneously detected by EIT, TRACE
$\lambda$195 and RHESSI (Lin et al. 2002). This flare occurred in the
active region AR0486 located at S18 E18. Fig.~\ref{goes} displays the
GOES-12 plot for this flare. The flare started around 10:20 UT with a
peak at around 10:45 UT and ended at around 11:00 UT.

\begin{figure}[!h]
\centering
\includegraphics[width=0.4\textwidth]{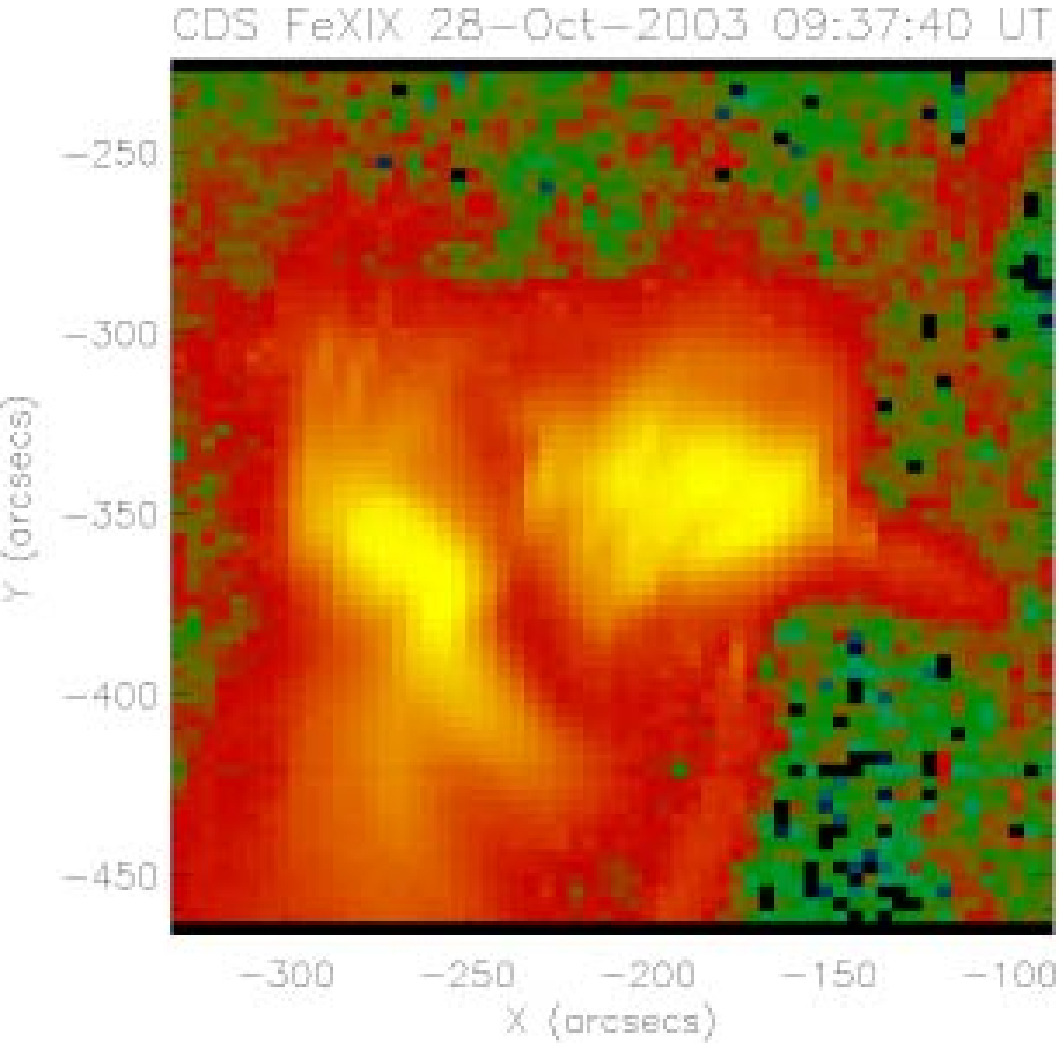}
\vspace{-0.9cm}

\includegraphics[width=0.4\textwidth]{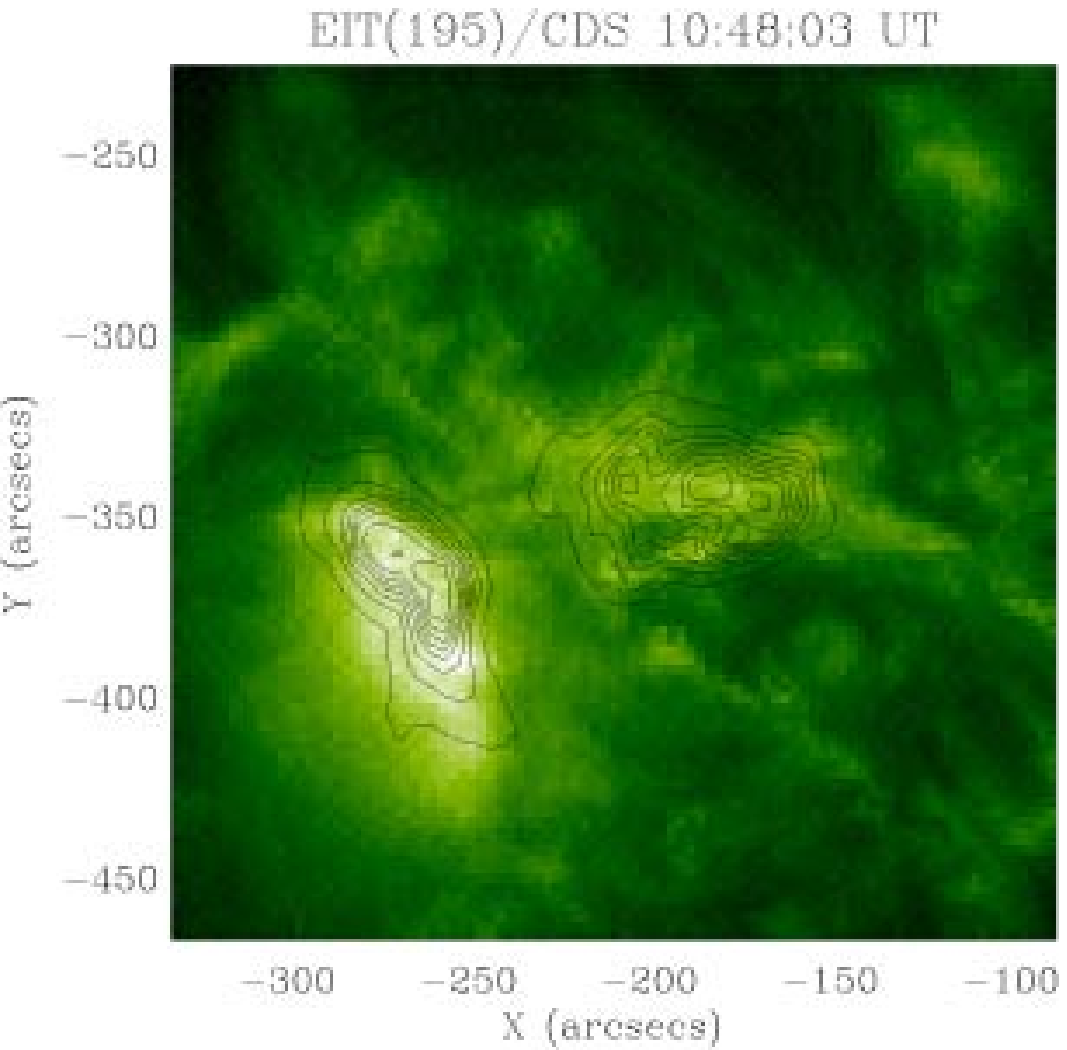}
\vspace{-0.9cm}

\includegraphics[width=0.4\textwidth]{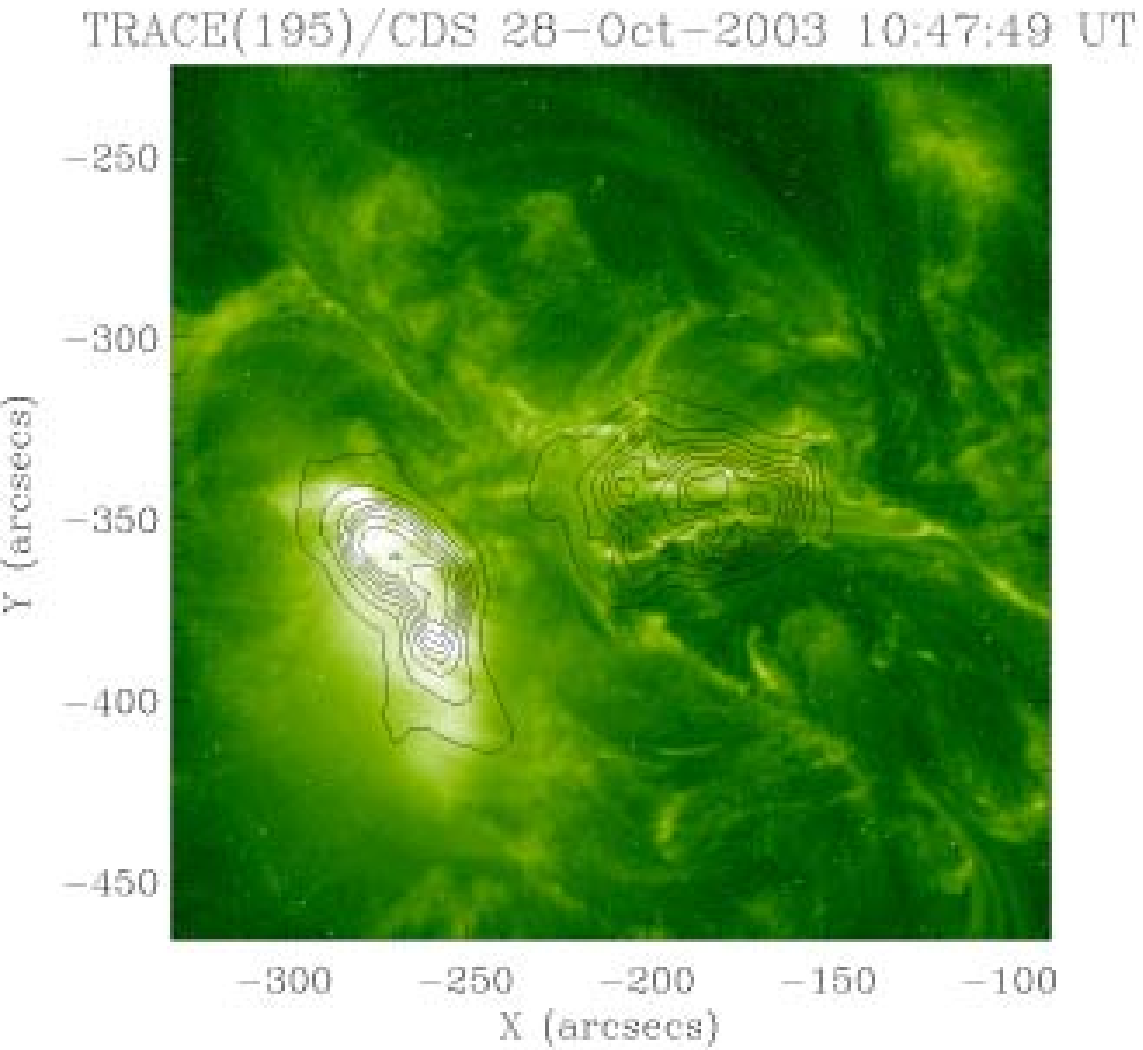}
\vspace{-0.9cm}

\includegraphics[width=0.4\textwidth]{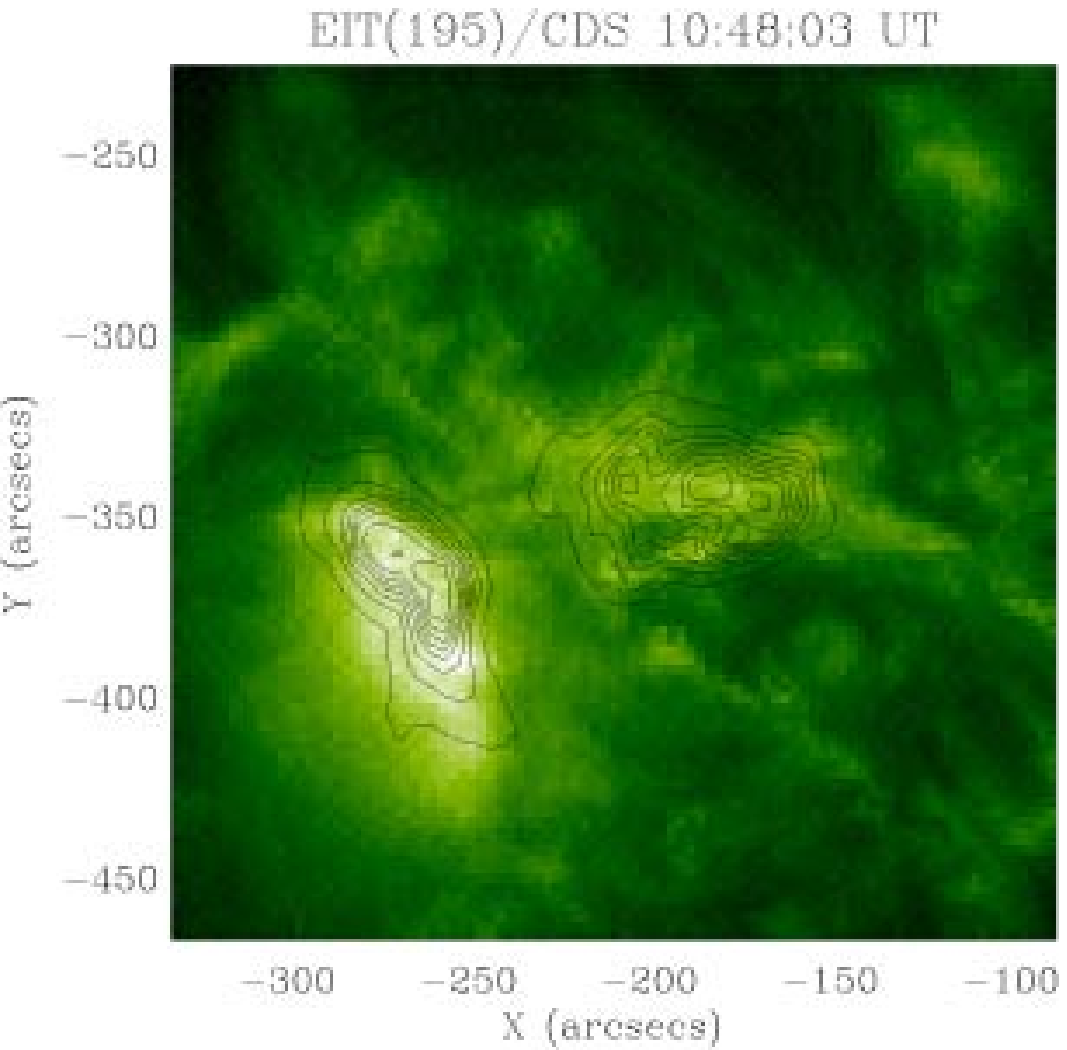}
\caption{A CDS image (top panel) in the Fe {\rm XIX} ($\lambda$592.2,
Log T=7.0) line. The time 09:37:40 UT in the title is the start time
of the raster. Images taken by EIT (second panel) and TRACE (third
panel) respectively at $\lambda$195 at 10:48:03 UT. The intensity
contours from the CDS image (in the top panel) obtained for Fe XIX
line are over plotted.  The EIT images are overlayed with RHESSI
contours in the 12-25 keV channel (bottom panel).\label{maps}}
\end{figure}

Fig.~\ref{maps} displays the images recorded by CDS in Fe {\rm XIX}
line, EIT and TRACE $\lambda$195 with overlayed Fe {\rm XIX} and
RHESSI intensity contours. The RHESSI image was obtained at 10:33 UT
by integrating the counts for about 20 sec. Note that the flare at
location (-280, -380) arcsec is of relevance in this paper as this was
observed simultaneously by all instruments.  The CDS raster scan
started at 09:37:40 UT (Solar X=-90, see Fig.~\ref{maps}) and it
observed the peak flare emission (Solar X=-250:-300) between 10:33 and
10:47 UT i.e., around the peak of the flare.

We have examined the sequence of TRACE (cadence: $\approx$ 1 min) and
EIT (cadence: 12 mins) images and found that the flare region varied
in intensity by less than 20\% during the CDS scan, as expected when
considering the GOES light curve.

\begin{figure}
\centering
\includegraphics[width=0.355\textwidth, angle=90]{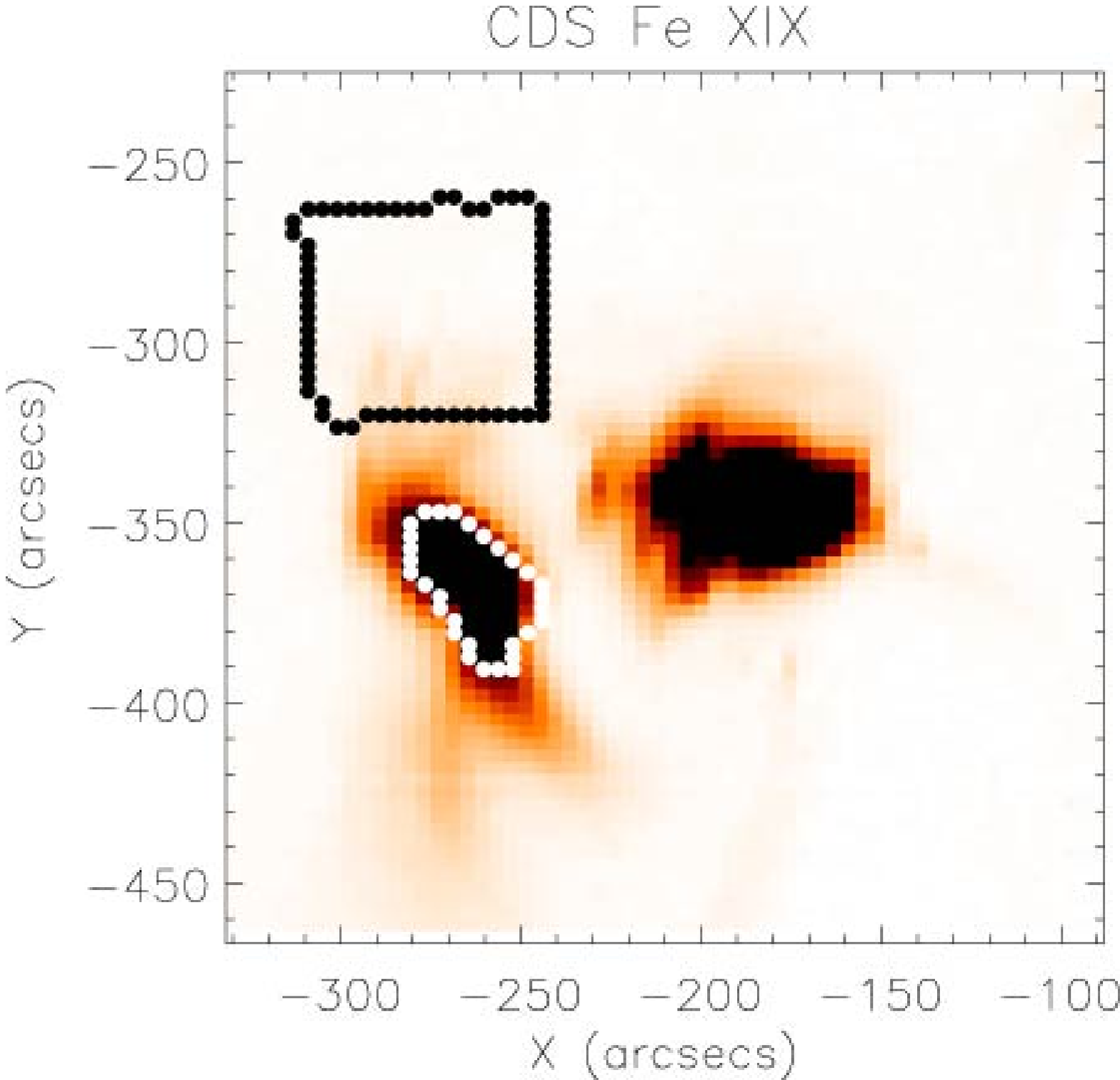}
\caption{CDS negative intensity image in Fe {\rm XIX} showing the two
regions selected for further analysis. White contour outlines the
flare region and the black outlines the outer region.
\label{region}}
\end{figure}
\begin{figure*}
\centering
\includegraphics[width=0.4\textwidth]{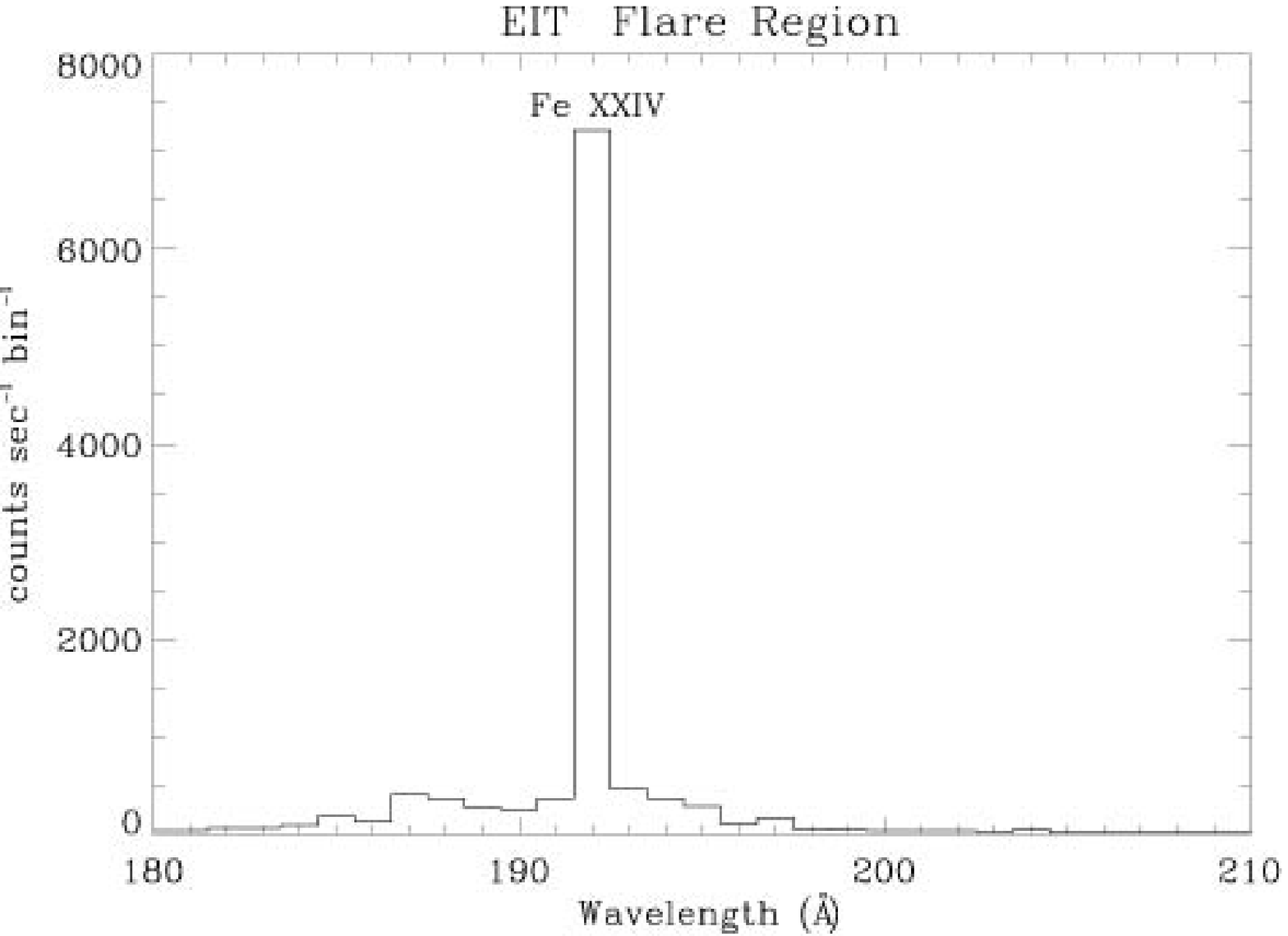}
\includegraphics[width=0.4\textwidth]{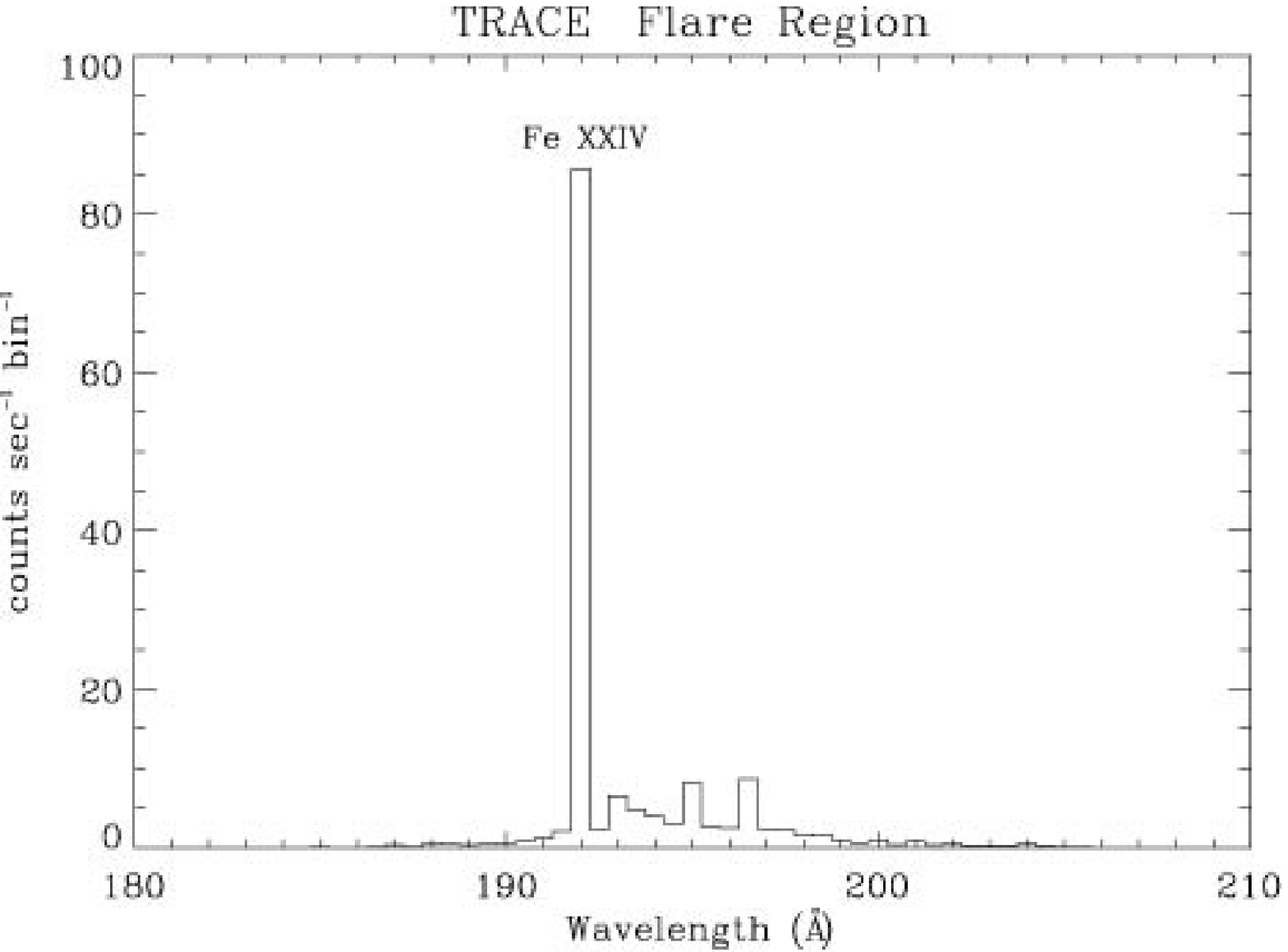}
\includegraphics[width=0.4\textwidth]{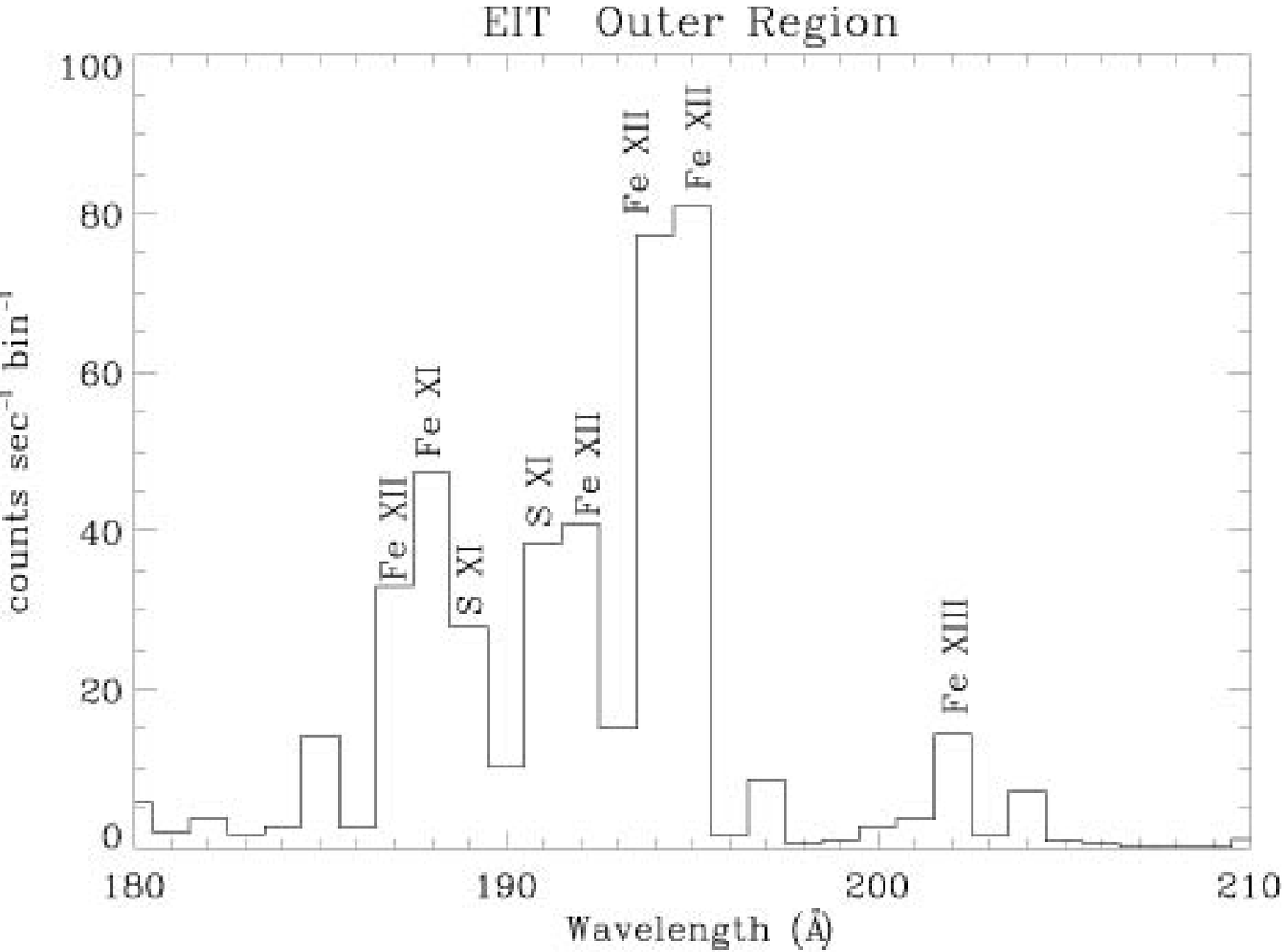}
\includegraphics[width=0.4\textwidth]{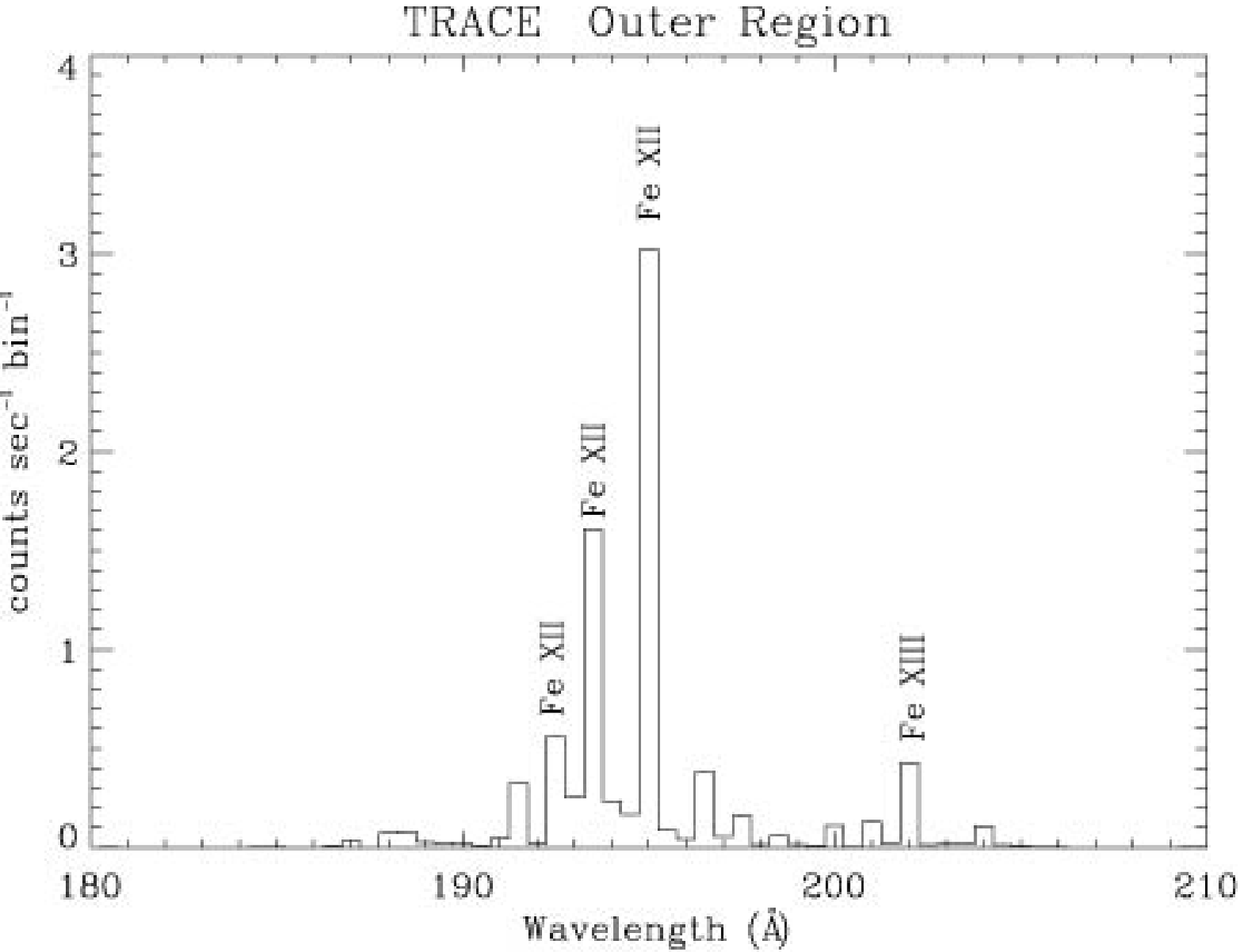}
\caption{CHIANTI synthetic spectra for the EIT (left hand side) and
TRACE (right hand side) $\lambda$195
channels in flare region (top panels) during a M8-class flare and
outer region (bottom panels).\label{spectrum}}
\end{figure*}

The standard routines provided in solar software (SSW) were used for
processing the EIT and TRACE images. The CDS data processing was
performed using routines and calibration described in Del Zanna et
al. 2001. In order to locate exactly the same region we coaligned the
images obtained by CDS, EIT and TRACE.  The TRACE image taken at 10:47
UT was first coaligned with the EIT images and then the CDS intensity
contours were overplotted.  As can be seen from Fig.~\ref{maps}, Fe
{\rm XIX} and RHESSI intensity contours match extremely well the peak
emission in the EIT and TRACE images.

\section{Forward and inverse modelling}
The intensity of an optically thin emission line can be written as 

\begin{equation}
 I = A(Z)\int_{Te}G(T_{e},N_{e})DEM(T_{e}) dT_{e}
\end{equation}

Where, A(Z) is the elemental abundance, $T_{e}$ is the electron
temperature, and $N_{e}$ is the electron density, G(Te, Ne) is the
contribution function which contains all the relevant atomic physics
coefficients. DEM($T_{e}$) is is defined as DEM($T_{e}$)=$Ne^2$(dh/d$T_{e}$), 
where dh is an element of column height along the line-of-sight.

\begin{table}
\begin{center}
\caption{Intensities in the flare region from the CDS NIS. Note that
intensity of Fe {\rm XXV} is deduced from RHESSI observations. In the
table from left to right: the ion, the observed wavelength, the
maximum formation temperature, the observed and predicted intensities
(phot $cm^{-2} sr^{-1} s^{-1}$) and the ratio of the latter
two.\label{lines_in}}
\begin{tabular}{lccrrc}
\hline
$Ion$ & ${\lambda}_{obs}$ &  $log T_{max}$ & $I_{obs}$ & $I_{dem}$ & $Ratio$\\
      & $({\AA})$           &  $(K)$         &\\
\hline
Ne {\rm IV}     &543.9 &      5.3   &   381.4  &    301.8 &    0.79
\\
O {\rm IV}      &553.3 &      5.3   &  1023.1  &   1029.8 &    1.01
\\
O {\rm V}       &629.7 &      5.4   &  6507.1  &   5254.3&     0.81
\\
Fe {\rm XI}     &369.2 &      6.1   &   159.8  &     91.5&     0.57
\\
Fe {\rm XII}    &364.5 &      6.1   &   813.9  &    607.9&     0.75
\\
Fe {\rm XIII}   &348.2 &      6.2   &   266.9  &    255.9&     0.96
\\
Fe {\rm XIV}    &334.2 &      6.2   &  1199.4  &   1245.9&     1.04
\\
Fe {\rm XVI}    &335.3 &      6.3   & 27134.7  &  25142.0&     0.93
\\
Mg {\rm X}      &624.9 &      6.0   &  3339.3  &   4230.2&     1.27
\\
Fe {\rm XIX}    &592.2 &      7.0   &  2368.7  &   2354.8&     0.99
\\
Fe {\rm XXV}    &1.833   &      8.1   &418305.1  & 286297.1 &    0.68
\\
\hline
\end{tabular}
\end{center}
\end{table}
\begin{table}
\begin{center}
\caption{Same as Table~\ref{lines_in} but for the outer
region.\label{lines_out}}
\begin{tabular}{lccrrc}
\hline
$Ion$ & ${\lambda}_{obs}$ &  $log T_{max}$ & $I_{obs}$ & $I_{dem}$ &
$Ratio$\\
      & $({\AA})$           &  $(K)$         &\\
\hline
O {\rm IV}    & 554.1 &  5.3  &       102.0   &  125.2   &    1.23
\\
O {\rm V}     & 629.7 &  5.4  &       1067.8  &  346.0   &    0.32
\\
Fe {\rm XI}   & 369.2 &  6.1  &       38.5    &  29.5    &    0.77
\\
Mg {\rm IX}   & 368.1 &  6.0  &       441.4   &  430.1   &    0.97
\\
Fe {\rm XII}  & 364.5 &  6.1  &       247.8   &  304.9   &    1.23
\\
Fe {\rm XIV}  & 334.2 &  6.2  &       401.7   &  717.7   &    1.79
\\
Fe {\rm XVI}  & 335.3 &  6.3  &       4022.7  &  3404.5  &    0.85
\\
\hline
\end{tabular}
\end{center}
\end{table}
\begin{table}
\begin{center}
\caption{Observed and predicted average count rates (DN $s^{-1} pixel^{-1}$) for the regions
marked in Fig.~\ref{region}.\label{counts}}
\begin{tabular}{ccccc}
\hline &\multicolumn{2}{c}{Flare Region}&\multicolumn{2}{c}{Outer
Region}\\
\cline{2-5}
&Observed & Predicted & Observed & Predicted \\
\hline
EIT     &   6375.9 & 11035.66&  572.5  &447.37\\
\hline
TRACE   & 45.6 & 149.71& 4.2 &8.25\\
\hline
\end{tabular}
\end{center}
\end{table}

To calculate the DEM($T_e$) the maximum entropy method described by
Monsignori Fossi and Landini (1991) and implemented by Del Zanna
(1999) was used. In this method the DEM is assumed to be a series of
spline mesh points, covering the temperature range for which the
observational constraints are present.  We measured the DEM of the
flare region and an outer region shown in
Fig.~\ref{region}. Tables~\ref{lines_in} and ~\ref{lines_out}
represent the lines in the flare and outer regions respectively which
where used for DEM analysis. As can be seen from the table, we have
been able to predict the intensities to within 30\% using the derived
DEM. The theoretical intensities were calculated by interfacing DEMs
within CHIANTI using a constant electron density of $10^{11}$
$cm^{-3}$ for the flare region and $10^9$ $cm^{-3}$ for the outer
region, the ionization fraction of Arnaud \& Rothenflug 1985 and
photospheric abundances.  However, we note that the choice of
ionization fraction and abundances has a negligible effect on the
results.  The $\lambda$195 channels are dominated by Fe lines, which
are observed by CDS, hence the choice of abundance only changes the
DEM by a scaling factor and not the simulated spectra.  The
$\lambda$195 channels turn out to be dominated, with the exception of
the flare case, by ions that are observed by CDS, so in this case the
main uncertainty lies in the relative intensity of lines observed at
different wavelengths.

In order to obtain the spectra, theoretically calculated line
intensities were convolved with the effective areas of EIT
$\lambda$195 {\it clear} and TRACE $\lambda$195 {\it A0} filters. The
final spectra are shown in Fig.~\ref{spectrum}. The difference in the
spectral resolution evident from Fig.~\ref{spectrum} is due to
difference in instrument response function. The TRACE response
function is narrower than that of EIT and hence TRACE has a better
spectral resolution. Note that in calculating the synthetic spectra,
the continuum was included. From Fig.~\ref{spectrum} it can be seen
that the spectrum of the flare region (top panels) shows a strong peak
at $\lambda$192 due to Fe {\rm XXIV} (20 MK) emission for both EIT and
TRACE.  However the spectra of the outer region (bottom panels) are
dominated in TRACE by Fe {\rm XII} lines.  The differences between
TRACE and EIT responses are significant.  The EIT emission has a
substantial contribution from other lines such as Fe {\rm XII}
$\lambda$187, Fe {\rm XI} $\lambda$188.2 \& 188.3, S {\rm XI}
$\lambda$188.7 \& $\lambda$191.25, Fe {\rm XII} $\lambda$192.4 \&
$\lambda$193.5 and Fe {\rm XIII} $\lambda$202.1.

Finally the average count rates obtained from the synthetic spectra
were compared with actual count rates obtained using the processed
images.  Table~\ref{counts} provides the average count rates for EIT
and TRACE corresponding to the flare and outer regions. As can be seen
from Table~\ref{counts} we could predict the average count rate for
EIT within 25\% for the outer region. However, the predicted count
rate for the flare region is a factor of two higher. In the case of
TRACE, the predicted count rate is a factor of two higher in the
outer region and a factor of 3 higher in the flare region.

\section{Discussion}

Using a similar forward modelling technique for the quiet Sun, Brooks
and Warren (2006) were able to predict the count rates within 25\% for
both EIT and TRACE $\lambda$195 channel using data recorded in
1998. Their result matches well with the one obtained here for the
outer region in the case of EIT but shows significant differences for
the TRACE count rates. This could be due to the fact that the TRACE
detector efficiency has decreased with time (as in the case of EIT).
The current EIT analysis software applies a correction factor of about
7 for the dataset considered here, to take into account decreased
sensitivity.  This suggests that the TRACE calibration might need to
be revisited.  Considering that most emission is due to Fe {\rm XII},
we can rule out any problems with the atomic data (Del Zanna \& Mason
2005).

The question also arises that, why is there such a big difference in the
observed and predicted count rates in the flare region? There could be
at least two possible reasons. The first is that our DEM is over
estimated. This could be due to the lack of constraint between the CDS
Fe {\rm XIX}$\lambda$592 (8 MK) and the RHESSI Fe {\rm XXV}
$\lambda$1.8 emission, or to an instrument calibration problem.

The second is that there remains some uncertainties in Fe {\rm XXIV}
emissivities. We note that strong discrepancies (factors 2-5) are
common in other cooler Li-like ions (Del Zanna et al. 2002).  This
issue needs to be further explored, if EIT,TRACE (and soon STEREO,
SDO) $\lambda$195 images of flares are to be reliably used for plasma
diagnostics.

Despite some differences in the count rates for the flare region for
EIT and TRACE and in outer region for TRACE, we have clearly
demonstrated that the EIT and TRACE $\lambda$195 channels are
dominated by Fe {\rm XXIV} $\lambda$192 emission during strong flares.

Future instruments, such as the Extreme-UltraViolet Imaging
Spectrometer (EIS) aboard Hinode (Solar-B), will provide a better
temperature coverage for the coronal component of active regions and
flares. This will better constrain the predicted spectrum,
enabling us to use CHIANTI to forward model the observational signal
for various future missions.

\begin{acknowledgements}

{We acknowledge the support of PPARC. We thank the SOHO/CDS, SOHO/EIT,
TRACE, GOES and RHESSI team for providing the data. SoHO is an
international collaboration between ESA and NASA. CHIANTI is a
collaborative project involving the NRL (USA), RAL (UK), and the
following Universities: College London (UK), Cambridge (UK), George
Mason (USA), and Florence (Italy).  GDZ warmly thanks DAMTP for the
hospitality during this collaboration.}

\end{acknowledgements}

\end{document}